\documentclass{moriond}

\def\Journal#1#2#3#4{{#1} {\bf #2}, #3 (#4)}

\def\PRL{\em Phys. Rev. Lett.}
\def\PRD{{\em Phys. Rev.} D}

\def\be{\begin{equation}}
\def\ee{\end{equation}}
\def\bea{\begin{eqnarray}}
\def\eea{\end{eqnarray}}

\newcommand{\Photo}{}

\begin{document}
\vspace*{4cm}
\title{The DAMIC Experiment at SNOLAB }
\author{ Mariangela Settimo for the DAMIC Collaboration}

\address{SUBATECH, CNRS/IN2P3, Universit\'e de Nantes, IMT-Atlantique, Nantes, France}

\maketitle\abstracts{
The DAMIC (Dark Matter in CCDs) experiment at the SNOLAB underground laboratory uses fully depleted, high resistivity CCDs to search for
dark matter particles with masses below 10 GeV/c$^2$. 
An upgrade of the detector using an array of seven 16-Mpixel CCDs (40 g of mass) started operation in February 2017. 
The new results, obtained with the current detector configuration, will be presented. Future plans for DAMIC-M, with a total mass of 1kg and a ionization threshold of 2 electrons, will be discussed.
}

\section{Introduction}

The DAMIC (DArk Matter In CCD) experiment at SNOLAB~\cite{DAMIC} employs the bulk silicon of scientific-grade charge-coupled devices (CCDs) as target for the interactions of dark matter particles. 
The low pixel readout noise and leakage current make DAMIC extremely sensitive to ionization signals from the interaction of dark matter particles with nuclei or electrons in the silicon target. The low mass of the silicon nucleus provides good sensitivity to WIMPs with masses in the range 1-10 GeV/c$^{2}$, while the small band gap of silicon provides sensitivity to dark matter-electron interactions that deposit as little as 1.1 eV in the target. 

An R\&D program has been conducted since 2013 to prove the performance of the detector~\cite{DAMIC,compton,ionisation}, provide measurements of the background contamination~\cite{background} and demonstrate the potentiality for dark matter searches~\cite{WIMPlimits,HiddenPhotons} with only a few grams of detector mass.
A detector with a total mass of 40~g of mass is installed at the SNOLAB underground laboratory (Canada) since 2017, and is currently taking data. 

\section{Experimental setup}

In DAMIC, the sensitive detector is the silicon bulk of high-resistivity fully depleted CCDs. Each device is a 16 Mpixels CCD, with a pixel size of (15$\times$15)$\mu$m$^2$ and a thickness of 675~$\mu m$. Each CCD is epoxied onto a silicon backing, together with a flex cable that is wire bonded to the CCD and provides the voltage biases, clocks and video signals required for its operation. These components are supported by a copper frame to complete the CCD module. The modules are installed in a copper box that is cooled to 140 K inside a vacuum chamber. The box is shielded by 18~cm of lead to attenuate external $\gamma$ rays, with the innermost 2-inches made of ancient lead, and by 
 42~cm-thick polyethylene to moderate and absorb environmental neutrons. 

\begin{figure}
\centering
\includegraphics[width=0.35\linewidth]{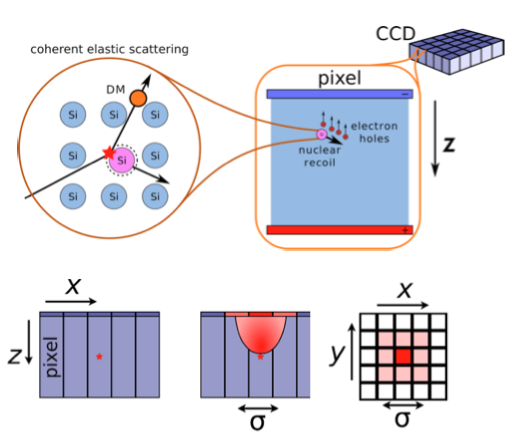}
\hspace{10mm}
\includegraphics[width=0.3\linewidth]{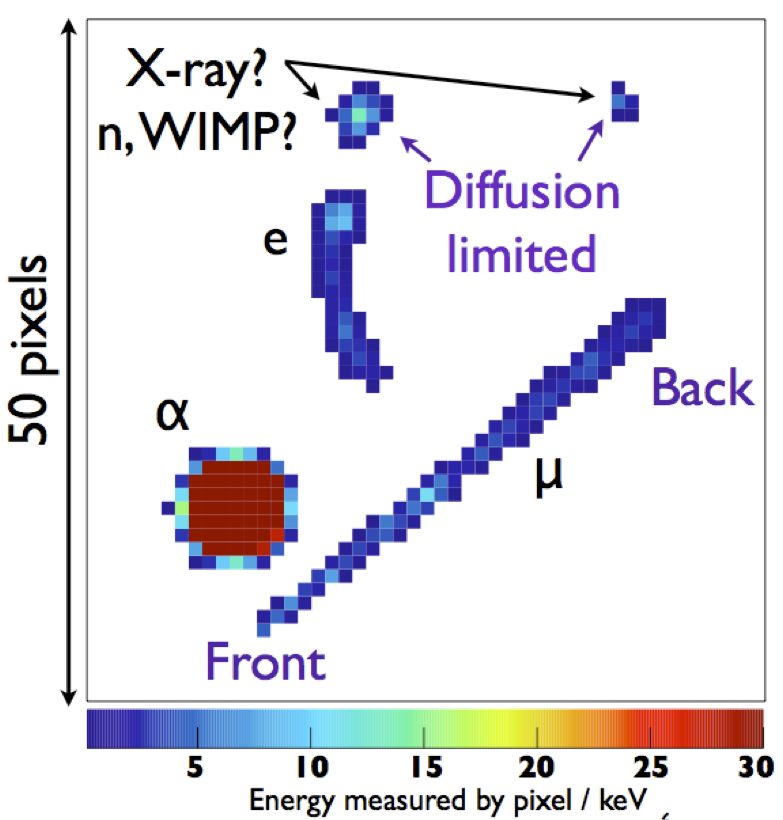}
\caption[]{Left: Representation of the WIMP-nucleon scattering and of the charge diffusion by a point-like ionization event in the CCD bulk. The x-y coordinates give the position in the CCD whereas the lateral spread positively correlates to the depth of the energy deposit. The diffusion model has been tested with data from radioactive sources and cosmogenic muons. Right: example of  detected tracks from different particle types.}
\label{fig:principle}
\end{figure}

When a DM particle scatter off a silicon nuclei (Fig.~\ref{fig:principle} left), the ionization charge is drifted along the direction of the electric field (z axis) and collected on the pixel array (x-y plane). Because of thermal motion, the charge diffuses transversely with a spatial variance that is proportional to the transit time (i.e., the depth of the interaction point). This allows for the reconstruction in three dimensions of the position of the energy deposit in the bulk of the device, and the identification of particle types based on the cluster pattern (Fig.~\ref{fig:principle} right). The exquisite spatial resolution and the 3D reconstruction are the basis for the rejection of background events produced by low energy gammas and electrons on the surface of the CCD and for the characterization of the radioactive background on the surface and in the bulk of the CCD~\cite{background}. 

The response of DAMIC CCDs to ionizing radiation has been measured with optical photons~\cite{DAMIC} for ionization signals down to 10 electrons, and with mono-energetic X and  $\gamma$-ray sources for energies in the range 0.5-60 keV$_{ee}$~\footnote{eV$_{ee}$ is the electron-equivalent energy scale relative to the ionization produced by recoiling electrons.}. 
Recoiling nuclei produce a smaller ionization signal (number of electron-hole pairs) than a recoiling electron of the same kinetic energy. 
We have measured the corresponding ionization efficiency in the energy range covered by DAMIC by comparing the observed and predicted nuclear-recoil energy spectra in a CCD from a low-energy $^{124}$Sb-$^{9}$Be photo-neutron source~\cite{ionisation} and a pulsed fast-neutron beam with a silicon drift detector~\cite{antonella}.
\begin{figure}
\centering 
\includegraphics[width=0.45\linewidth]{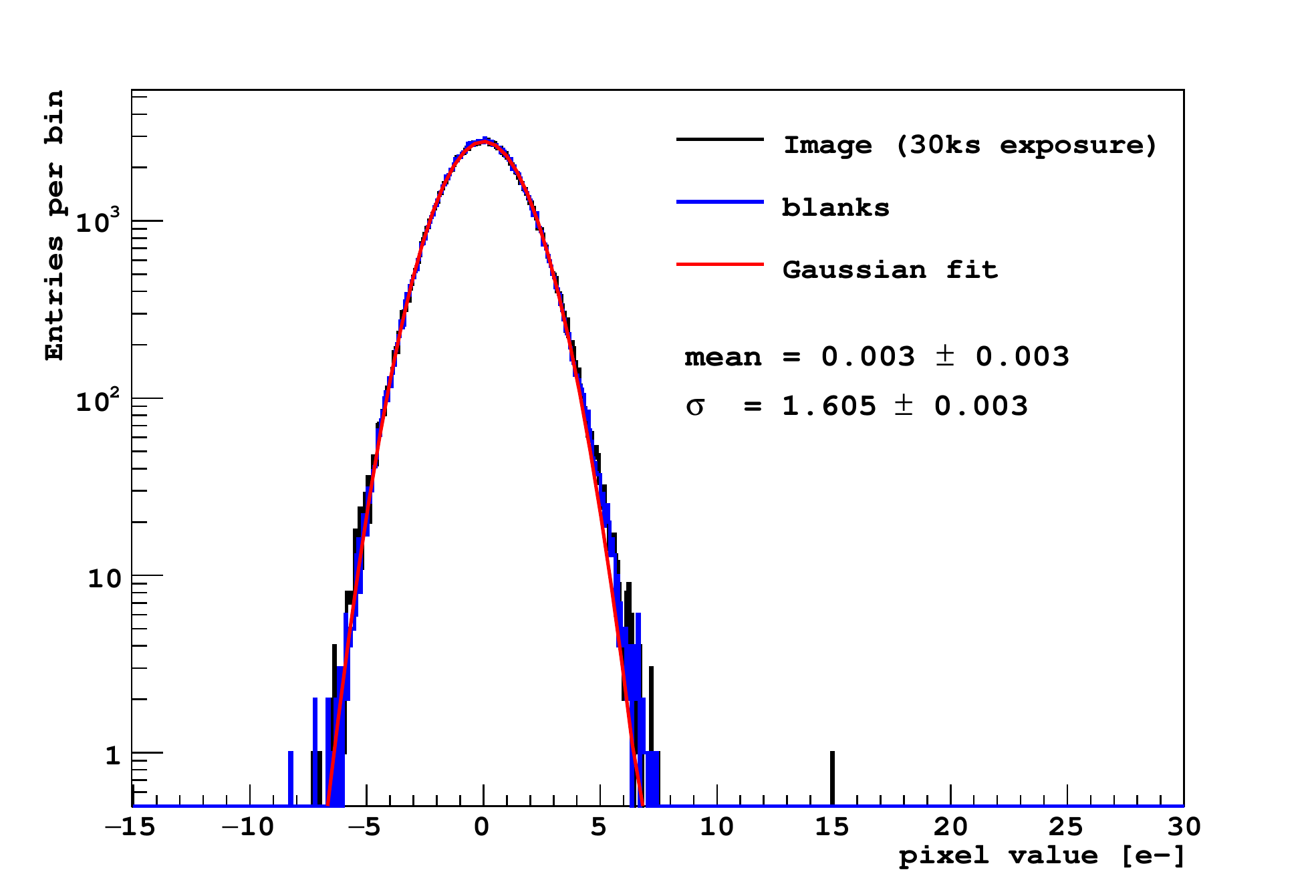}
\includegraphics[width=0.45\linewidth]{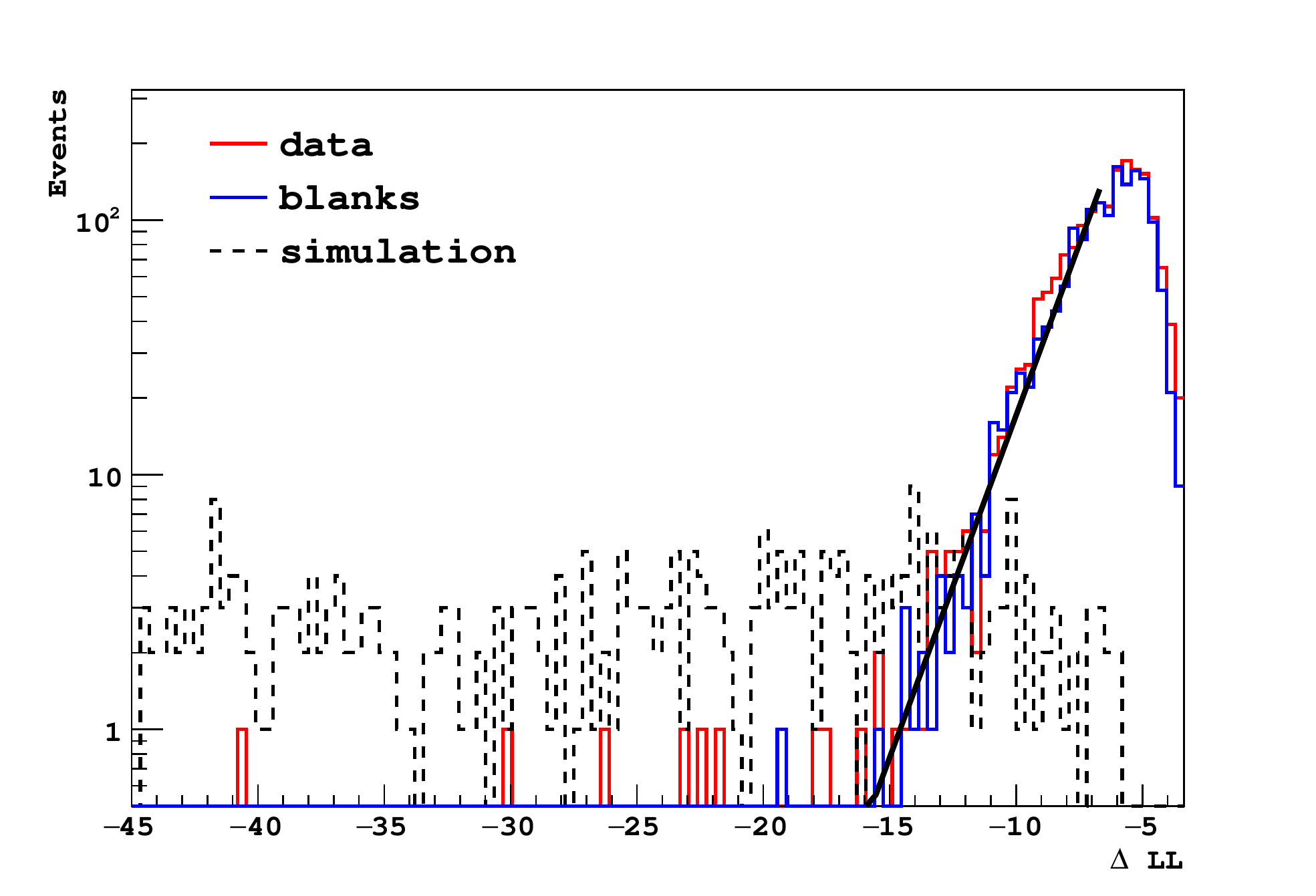}

\caption[]{Left: pixel value distribution in blanks images (white noise) and in 8 hours exposure (white noise, leakage current and signal) when operating at 140~K. 
Right: $\Delta_{LL}$ distributions for all clusters, in the blanks (blue), in simulations (black) and in data (red). }
\label{fig:analysis}
\end{figure}

\section{Current status and data analysis}
\begin{figure}
\includegraphics[width=0.49\linewidth]{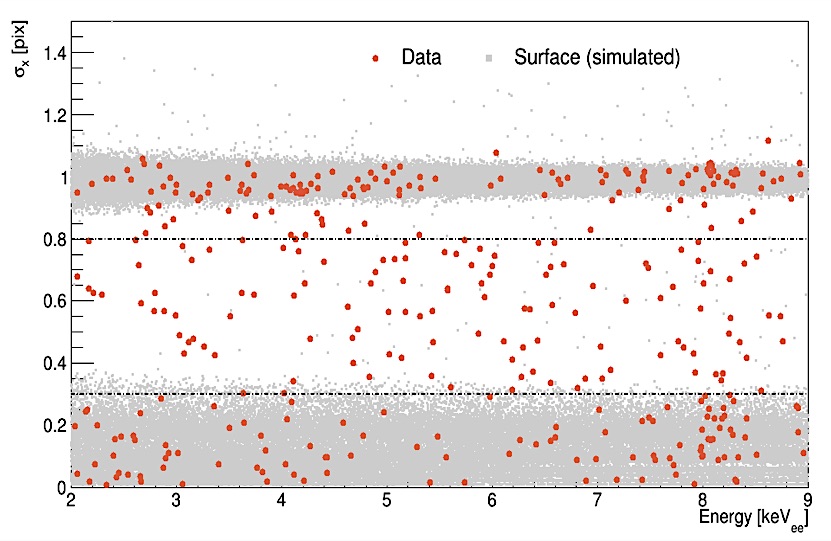}
\includegraphics[width=0.51\linewidth]{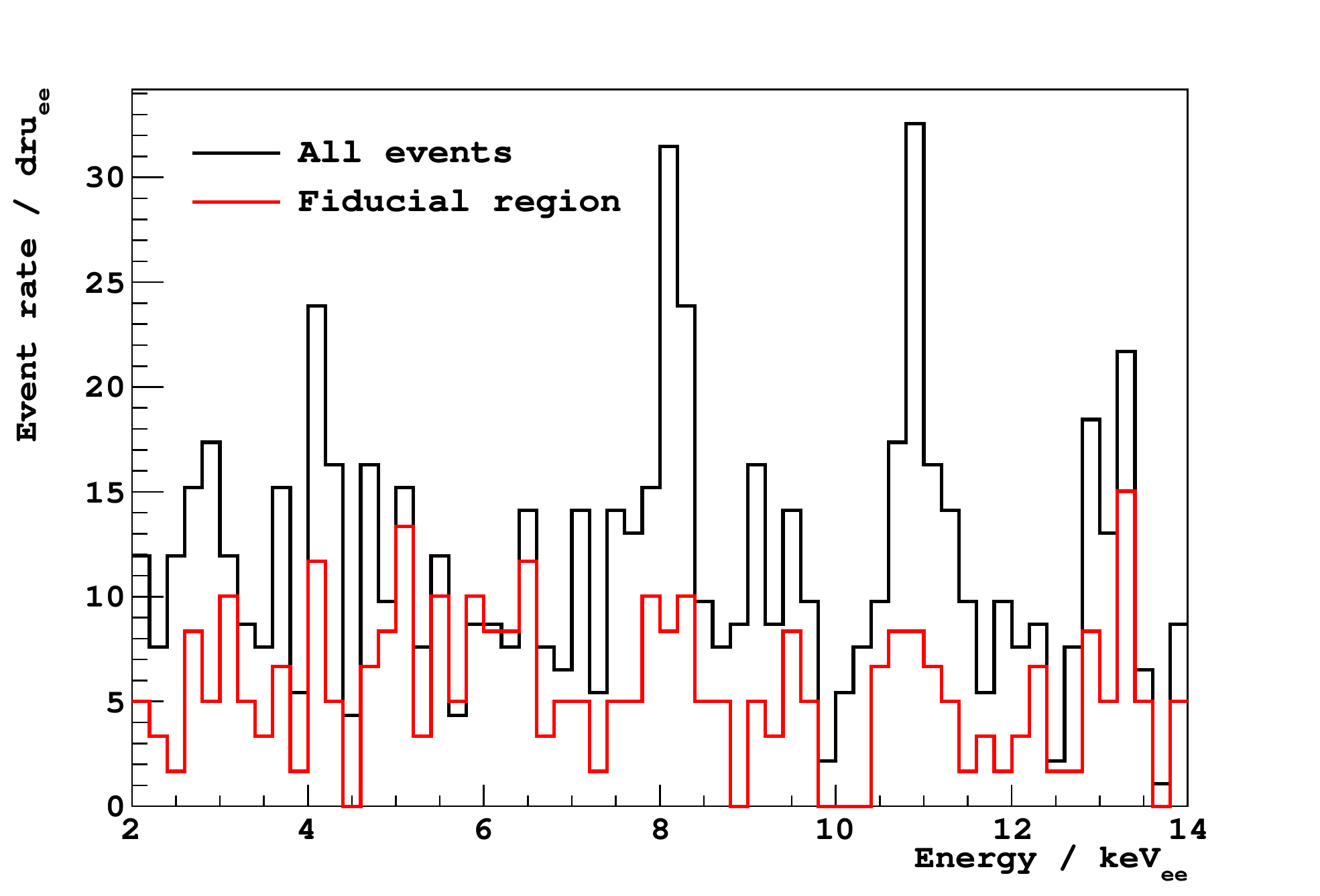}
\caption[]{Left: Lateral diffusion ($\sigma_{x}$) versus measured energy (E) of the clusters that pass the selection criteria.  Gray markers show the simulated energy deposits on the surface of the device. Red markers are the reconstructed events in data. The horizontal dashed lines delimits the fiducial region, rejecting events on the front ($\sigma>$ 0.8) and back ($\sigma<$0.3) of the CCD. Right: Spectrum of the reconstructed events at energy above 2 keV.}
\label{fig:spectrum}
\end{figure}
The current detector installation consists of 7 CCDs (one of them sandwiched between two blocks of ancient lead) for a total mass of 40~g. A data set corresponding to about  7.6~kg~day exposure, has been acquired for background studies and data characterization.  A second set of data (4.6 kg day exposure, as of March 2018) has been acquired for dark matter searches. In the first case, the charge collected by each pixel is read out individually, offering maximum spatial resolution. In the second data set, the charge collected by column segments of 100 pixels in height is read in a single measurement. As the charge from an ionization event is then distributed over a smaller number of measurements, the readout noise contribution in the determination of the energy of the cluster is reduced, improving the energy threshold. 

Images were acquired with 8 or 24 hour exposures, each followed by a zero-length exposure (``blank'') for noise and detector monitoring.  
Figure ~\ref{fig:analysis} (left) shows the pixel value distribution as observed in blanks images (white noise) and in 8 hours exposure (convolution of white noise, leakage current and signal). This distribution of the image is consistent with the one expected for a  white noise, and is well described by a Gaussian distribution with a standard deviation of $\sim 1.6$~electrons (about 6 eV$_{ee}$) and a leakage current as low as 10$^{-3}$~e$^{-}$/pix/day. 

The pre-processing of the images, with the pedestal and correlated noise subtraction is discribed in details in~\cite{WIMPlimits}. To select cluster events due to energy deposit in the CCD, we perform a scan of the image with a moving window: For every position of the window, we compute the likelihood $\mathcal{L}_n$ that the pixel values in the window are described by white noise and the likelihood $\mathcal{L}_G$ that the pixel values in the window are described by a Gaussian function plus the white noise. Large (or less negative) values of $\Delta LL = \mathcal{L}_n - \mathcal{L}_G$ are expected from noise events, as shown in Figure~\ref{fig:analysis},  for the blank images (blue), the simulated energy deposits (dashed black) and for real images (red) where events due to noise and energy depositions are both present. 
A selection on the value of dLL can be defined to efficiently reject clusters that arise from readout noise. 
To select a data set of events with reduced background contamination, surface events are rejected applying a fiducial region cut based on the diffusion parameter $\sigma$ (Figure~\ref{fig:spectrum}, left). Preliminary analysis of the current data suggests that an energy threshold of 50 eV$_{ee}$ can be achieved with an expected leakage of events from readout noise in the entire data set of only 0.01. 
The energy spectra of the reconstructed events after the $\Delta LL$  criterion (black) and after the fiducial region cut (red), is shown in Figure~\ref{fig:spectrum} (right), at energies above 2~keV. The peak at 8 keV is consistent with the expected copper fluorescence line and is removed when selecting only events in the bulk of the CCD. 
The background level between 0.5 keV and 14.5 keV is lowest for the CCD sandwiched in between the two ancient lead blocks, with a value of 2 DRU, while it is consistent with 5 DRU~\footnote{1 DRU = 1 event keV$^{-1}$ kg$^{-1}$ d$^{-1}$.} in all other CCDs. The analysis of the data below 2~keV is ongoing.  
Figure~\ref{fig:results} (left) shows the expected sensitivity of DAMIC to the spin-independent WIMPs-nucleus cross-section, with the current 4.6~kg~day exposure (dashed) and with a 13~kg~day exposure (pointed) expected by the end of 2018. With the current exposure, we can explore a large parameters phase-space of the signal-excess reported by the CMDS collaboration~\cite{CDMS} using the same nuclear target and an energy threshold that is improved by a order of magnitude compared to CDMS-II Si. 

\section{Future plans}
\begin{figure}
\centering
\includegraphics[width=0.51\linewidth]{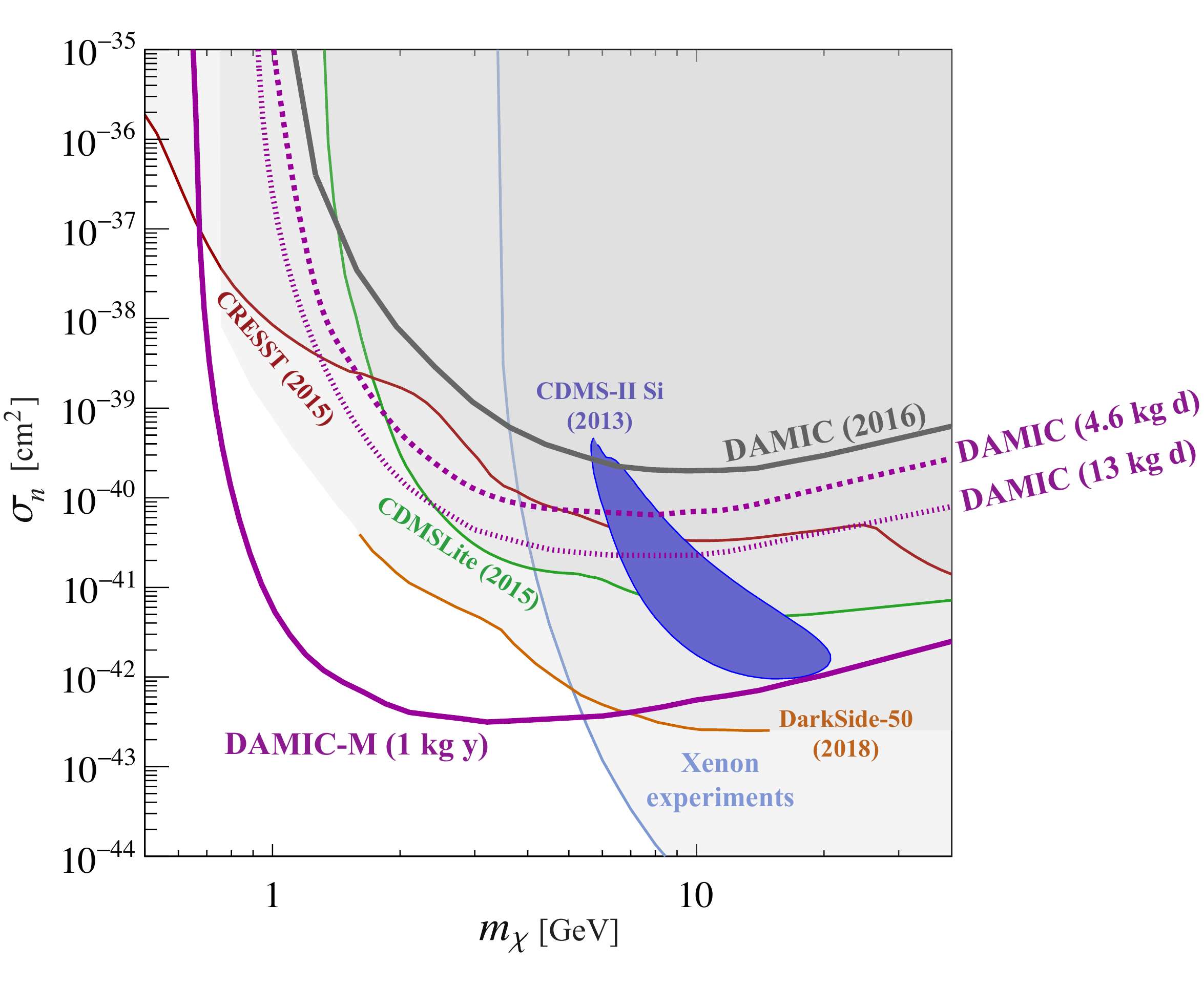}
\includegraphics[width=0.43\linewidth]{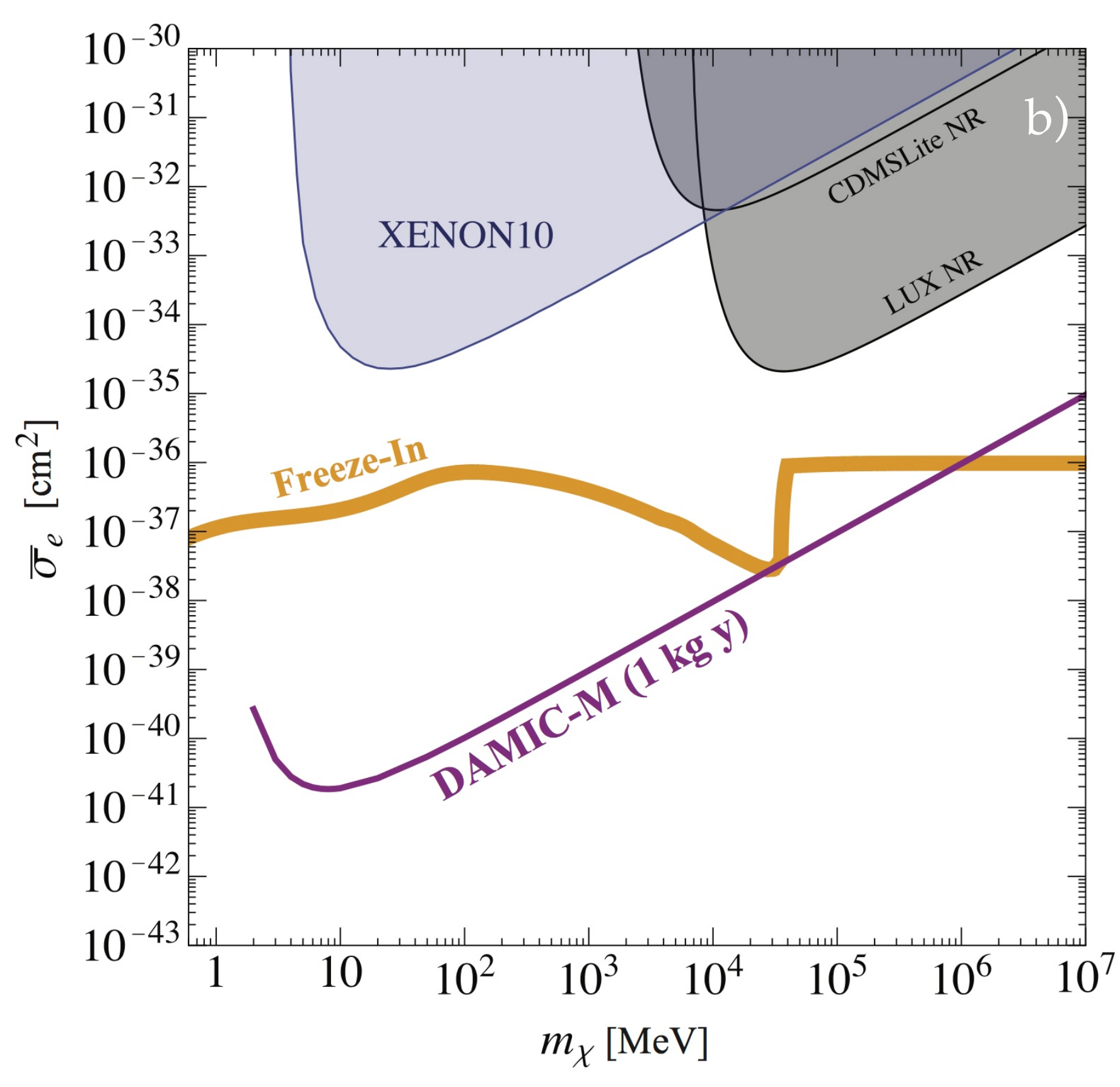}
\caption[]{Left: Expected sensitivity of DAMIC100 and DAMIC-M for WIMP-nucleon spin-independent scattering. Exclusion limits from other dark matter searches
are shown for comparison. Right: DM-electron cross section vs m$_{\chi}$ for a light mediator~\cite{science}.}
\label{fig:results}
\end{figure}
The next step in the DAMIC program is the construction of DAMIC-M at the Modane underground laboratory in France. The detector will employ new designed large CCDs (36~Mpixels size, 1~mm thickness and a mass of about 20~g each) and will reach a kg-size mass. 
DAMIC-M CCDs will adopt the``skipper'' readout stage with single electron resolution designed by Berkeley Lab and tested in~\cite{SENSEI}. When implemented on a DAMIC CCD with its demonstrated extremely low level of leakage current, will allow for a threshold of 2 or 3 electrons. 
Several improvements in the detector design, construction materials and CCD packaging are foreseen in order to decrease the background level to a fractions of a DRU. Under these conditions, DAMIC-M will be able to progress further in the search for low-energy dark matter particles, including the GeV-scale WIMPs (Fig.~\ref{fig:results} left, solid line), the hidden-photon, and to probe a large region of parameter space for dark matter particle in the ``hidden sectors'' (not directly coupling with the ordinary matters) and having masses from 1~MeV/c$^2$ to 1~GeV/c$^2$ (see for example Fig.~\ref{fig:results} right and Ref.~\cite{science}).

\end{document}